\documentclass[a4paper,11pt]{article}
\usepackage{pos}
\usepackage{tikz}
\usetikzlibrary{shapes,arrows,positioning,fit,backgrounds,calc}
\usepackage{caption}
\usepackage{subcaption}

\title{Sensitivity of Double Deeply Virtual Compton Scattering observables to GPDs}

\author*{J. S. Alvarado}
\author{M. Hoballah}
\author{E. Voutier}

\affiliation{Université Paris-Saclay, CNRS/IN2P3, IJCLab, 91405 Orsay, France}

\emailAdd{alvarado-galeano@ijclab.in2p3.fr}

\abstract{...}

\FullConference{31st International Workshop on Deep Inelastic Scattering (DIS2024)\\
 8–12 April 2024\\
Grenoble, France\\}

\abstract{Generalized Parton Distributions (GPDs) are multidimensonal structure functions that encode the information about the internal structure of hadrons. Using privileged channels such as Deeply Virtual Compton Scattering (DVCS) or Timelike Compton Scattering (TCS), it is possible to make direct measurements at points where the momentum fraction of the parton equals the respective scaling variable. Double Deeply Virtual Compton Scattering (DDVCS) is a not yet measured and promising channel for GPD studies as it allows to perform more general measurements at independent momentum fraction and scaling variable values. GPDs are extracted from Compton Form Factors which arise naturally in experimental observables from different combinations of beam and target configurations. In the context of the Continuous Electron Beam Accelerator Facility (CEBAF) and the Electron Ion Collider (EIC), we report the results of an exhaustive study of the DDVCS observables from polarized electron and positron beams directed to a polarized proton target. The study focuses on the sensitivity of the observables to the parton helicity conserving proton GPDs, particularly the consequences for GPDs measurements via DDVCS at CEBAF and EIC based on the VGG and GK19 model predictions.}

\begin{document}
\maketitle
\section{Introduction}
\begin{figure}[ht]
    \centering
    \begin{subfigure}[b]{0.3\textwidth}
        \centering
        \includegraphics[width=0.7\textwidth]{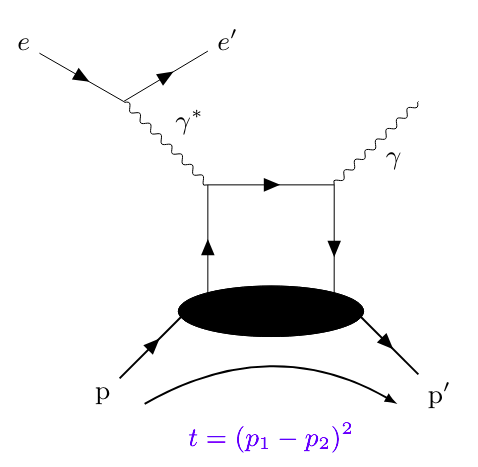}
    \end{subfigure}
    \begin{subfigure}[b]{0.3\textwidth}
        \centering
        \includegraphics[width=0.65\textwidth]{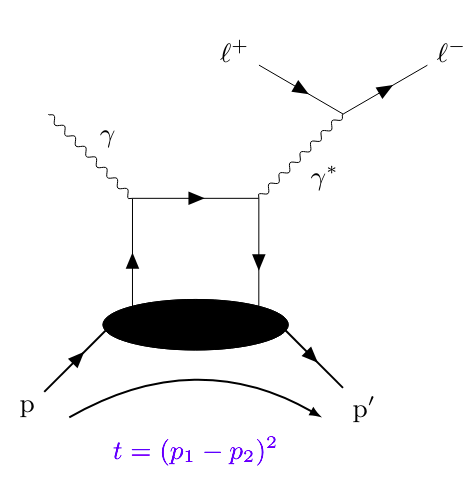}
    \end{subfigure}
    \begin{subfigure}[b]{0.3\textwidth}
        \centering
        \includegraphics[width=0.8\textwidth]{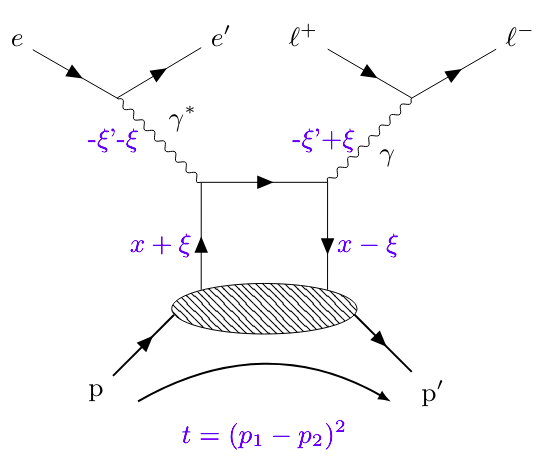}
    \end{subfigure}
    \caption{Handbag diagrams of DVCS (left), TCS (middle) and DDVCS (right).}
    \label{Diagrams}
\end{figure}

Generalized Parton Distributions (GPDs) are structure functions that correlate the transverse position and longitudinal momentum of quarks inside the nucleon. They encode spatial and momentum densities that can provide a 3D picture of the nucleon \cite{GPD3}, information about the mechanical properties of the nucleon \cite{GPD4} and information about the angular momentum contributions through Ji's sum rule \cite{GPD5}. Currently, we access GPDs through processes like Deeply Virtual Compton Scattering (DVCS), Time-Like Compton scattering (TCS) or Double DVCS (DDVCS) \cite{DDVCS1,DDVCS2} (Fig. \ref{Diagrams}) that are produced with high-energy lepton beams directed to a nuclear target. However, GPD information is not directly measurable. They enter the cross-section through Compton Form Factors (CFFs) given at leading order and leading twist by
\begin{align}
    \mathcal{F}(x\equiv\xi',\xi , t) &=\mathcal{P}\int_{-1}^1dx'~F(x',\xi,t)\bigg[\frac{1}{x'-\xi'}\pm\frac{1}{x'+\xi'}\bigg] -i\pi\big[F(\xi',\xi,t)-F(-\xi',\xi,t)\big], \label{CFF}
\end{align}
\noindent
where $F=H,E,\widetilde{H}, \widetilde{E}$ stands for the four existing chiral-even GPDs of a $1/2$ spin particle, $\xi$ is the skewness parameter, $t$ the momentum transfer of the nucleon, $x\equiv \xi'$ the fraction of the nucleon longitudinal momentum carried by the struck quark. The $\xi$ and $\xi'$ parameters can be written in terms of the photon virtualities and the Bjorken scaling variable $x_{B}$ as:
\begin{align}
    \xi'&=\frac{Q^{2} - Q^{\prime 2}}{2Q^{2}/x_{B} - Q^{2} - Q^{\prime 2} + t} &
    \xi&=\frac{Q^{2} + Q^{\prime 2}}{2Q^{2}/x_{B} - Q^{2} - Q^{\prime 2} + t}.
\end{align}
Inside CFFs, GPD information is encoded in a convolution integral on the real part and a direct GPD measurement on the imaginary part. For DVCS/TCS case, the GPD measurement is limited to the diagonal $x=\xi'=\pm \xi$.  The DDVCS case generalizes this feature and has the main benefit of having independent $\xi$ and $\xi'$ parameters thanks to the independent virtualities of the time-like ($Q^{2}$) and space-like ($Q^{\prime 2}$) photons \cite{Victor}, allowing then a general scan of GPDs.
\section{Experimental configuration and observables}
\noindent
Compared to DVCS/TCS, the DDVCS process has a much smaller cross-section and its identification requires a muon pair in the final state. The latter because for an electron/positron pair, one cannot guarantee the distinction of the scattered electron and the decay electron. As a result, measuring the DDVCS process requires a high luminosity and a large acceptance detector together with a dedicated muon detector \cite{CLAS12,SoLID}.

In spite of the experimental challenges for DDVCS, we are interested here in the GPD dependence of the experimental observables arising from polarized $e^{\mp}$ beams and targets, and their sensitivity to models. Namely the Beam Spin Asymmetry ($A_{LU}$), Target Spin Asymmetry ($A_{UL}$), Double Spin Asymmetry ($A_{LL}$) and the unpolarized Beam Charge Asymmetry ($A_{UU}^{C}$). 

We compute the prediction of these observables within the VGG (see \cite{VGG} for a review), GK19 \cite{GK1,GK2}, KM10,  KM15 \cite{KM} and AFKM12 \cite{AFKM} models. The GK model is taken from PARTONS \cite{PARTONS} while the KM models from Gepard \cite{Gepard}. However, there is no support for DDVCS computations on Gepard, so we extended CFF computations at $\xi'\neq \xi$ dependence. The $(\xi, \xi')$ dependence of the sea quark contribution is obtained from the Mellin-Barnes representation on \cite{KM}. For the valence quark contribution, it was implemented a custom Double Distribution able to reproduce the $\xi=\xi'$ result on \cite{GPDVal}. The observables were computed on a $(Q^{2}, Q^{\prime 2})$ grid at fixed $t$ and $x_{B}$ values. In general, we look for regions of important model sensitivity and we study the feasibility of such measurements for the Jefferson (JLab) CLAS12 and the Electron Ion Collider (EIC) scenarios.

\subsection{Sensitivity at JLab kinematics}
\begin{figure}[ht]
    \centering
    \includegraphics[width=0.7\textwidth]{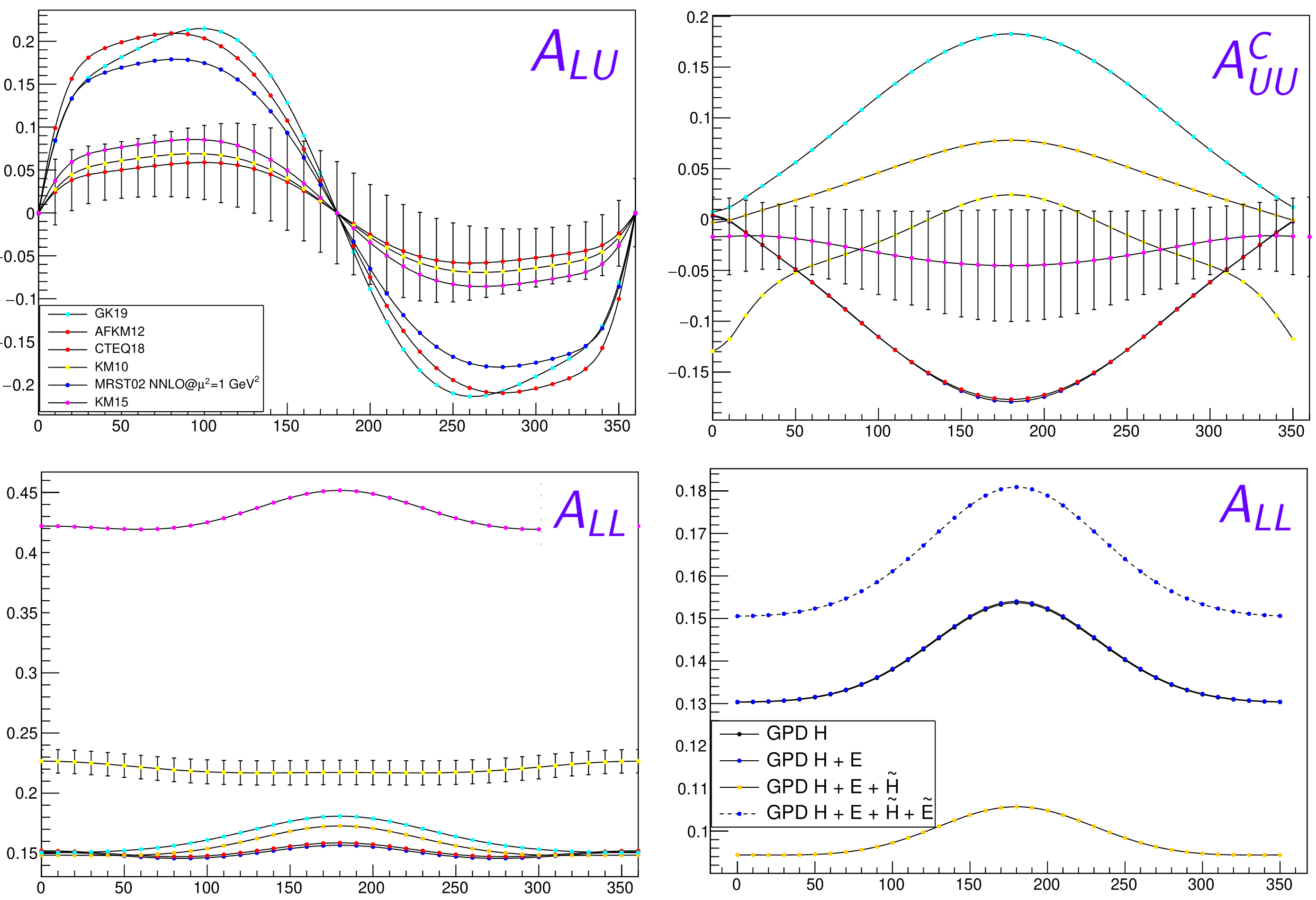}
    \caption{$A_{LU}$ (top left) and $A_{UU}^{C}$ (top right) at $x_{B}=0.15$, $t=-0.15$ GeV$^{2}$, $Q^{2}=2.77$ GeV$^{2}$, $Q^{\prime 2}=1.0$ GeV$^{2}$, $A_{LL}$ (bottom left) model predictions and $A_{LL}$ GPD dependence (bottom right) at $x_{B}=0.07$, $t=-0.15$ GeV$^{2}$, $Q^{2}=0.8$ GeV$^{2}$, $Q^{\prime 2}=2.4$ at JLab kinematics.}
    \label{JLab1}
\end{figure}

The CLAS12 detector at JLab currently supports a luminosity of $10^{35}$ cm$^{-2}$s$^{-1}$. Although suitable for DVCS measurements, a 100 times larger luminosity is required to enable DDVCS measurements. Considering the latter luminosity as nominal, a combined reconstruction and acceptance efficiency of $5\%$ and $100$ days of beam time, we present in Fig. \ref{JLab1} benchmark model predictions for $A_{LU}$, $A_{UU}^{C}$ and $A_{LL}$ at $t=-0.15$ GeV$^{2}$. The $(Q^{2},Q^{\prime 2})$ scan was done in steps of $\Delta Q^{2} = 0.4$ GeV$^{2}$ and $\Delta Q^{\prime 2}= (Q^{\prime 2}_{max} - 4m_{\mu}^{2})/10$. On the one hand, $A_{UL}$ is GPD $\widetilde{H}$ dominated but unmeasurable within 100 days of beam time and thus it was omitted. On the other hand, $A_{LU}$ and $A_{UU}^{C}$ are GPD $H$ dominated and have a strong model dependence in the amplitude and sign. Moreover, the error bars support the feasibility of these measurements within $100$ days of beam time and has the potential to set bounds on models. This last feature can be better exploited with $A_{LL}$ measurements where the model dependence is stronger and the GPD dependence is more complex. 

\subsection{Sensitivity at EIC kinematics}
\begin{figure}[ht]
    \centering
    \includegraphics[scale=0.31]{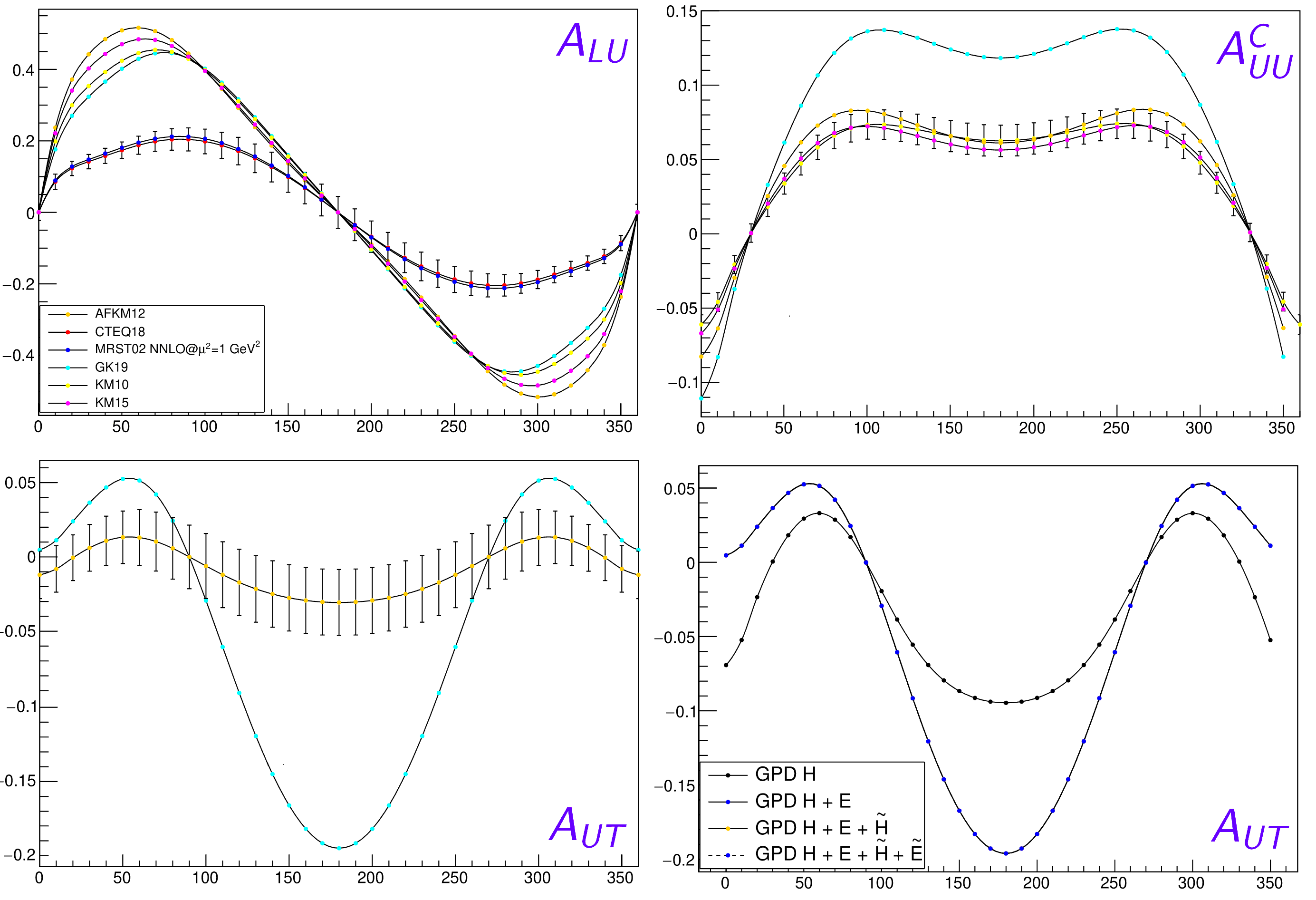}
    \caption{$A_{LU}$ (top left), $A_{UU}^{C}$ (top right), $A_{UT}$ (bottom left) model predictions and $A_{UT}$ GPD dependence (top right) at $x_{B}=10^{-4}$, $t=-0.15$ GeV$^{2}$, $Q^{2}=1.5$ GeV$^{2}$, $Q^{\prime 2}=0.34$ GeV$^{2}$ at EIC kinematics.}
    \label{EIC}
\end{figure}

In the EIC scenario, the luminosity is asummed to be $\mathcal{L}=10$ fb$^{-1}$ year$^{-1} \sim 10^{33} - 10^{34}$ cm$^{-2}$s$^{-1}$ with a center of mass energy reaching top values of $140$ GeV. Thus, the EIC would allow us to study the nucleon structure at small values of $x_{B}$ over a large ($Q^{2}, Q^{\prime 2}$) range. However, the cross section decreases quickly with $Q^{2}$ and $Q^{\prime 2}$, so the phase space exploration is performed on a kinematic region with a cross-section large enough to compensate the smaller luminosity of the EIC configuration. Such region is defined by $0.5\;\text{GeV}^{2}<Q^{2}<3\;\text{GeV}^{2}$ and $4m_{\mu}^{2}<Q^{\prime 2}<3\;\text{GeV}^{2}$ at $x_{B}=10^{-3}, 10^{-4}$ and $t=-0.15$ GeV$^{2}$. The $(Q^{2},Q^{\prime 2})$ exploration is done in steps of $0.5$ GeV$^2$ in $Q^{2}$ and $0.3$ GeV$^2$ in $Q^{\prime 2}$. As a result, we found that the longitudinally target polarized observables $A_{UL}$ and $A_{LL}$ are not suitable for measurements as their amplitudes are smaller than $1\%$ in the investigated region, in agreement with previous model predictions \cite{SY}. For unpolarized target observables, a benchmark set of predictions considering one effective year of data taking and a combined $5\%$ acceptance and reconstruction efficiency is shown in Fig. \ref{EIC}. We can observe an important model sensitivity that discriminates the VGG model with the $A_{LU}$ measurement and the GK19 model with the $A_{UU}^{C}$ measurement.

In addition to longitudinally polarized target observables, the EIC case opens the possibility of measuring transversely polarized target observables. Thus, we performed an exploration of the $A_{UT}$ sensitivity over the same phase space. The benchmark model prediction together with the GPD dependence in Fig. \ref{EIC}, shows a strong GPD $E$ dependence. In general, there is a lack of experimental data to constrain $E$. As a consequence, most of the GPD models do not offer support for $E$ leaving us with the GK19 and AFKM12 models only. Even though, we observe an important model sensitivity of such observable which together with the GPD $E$ dependence enhances the crucial role of $A_{UT}$ measurements in the near future for DDVCS and GPD studies.

All in all, DDVCS is a promising channel for GPD studies as it allows their exploration at independent $\xi$ and $\xi'$ values. Assuming polarized $e^{\mp}$ beam and targets, muon detection and a combined acceptance and reconstruction efficiency of $5\%$, we find that $A_{LU}$, $A_{LL}$, $A_{UU}^{C}$ are measurable within 100 days of beam time with an upgraded version of the CLAS12 detector at JLab. Likewise, we find that $A_{LU}$, $A_{UU}^{C}$ and $A_{UT}$ are measurable within a year of data taking at EIC kinematics. In both cases, the observables show an important model sensitivity with the potential to set bounds on models, a complex GPD dependence for $A_{LL}$ and a strong GPD $E$ dependence on the transverse polarized target case.

\noindent
\textbf{Acknowledgements}: This work was supported by the Programme blanc of the physics Graduate School of the PHENIICS doctoral school [Contract No. D22-ET11] and the french Centre National de la Recherche Scientifique (CNRS).

\end{document}